\documentclass[pre,aps,nofootinbib,twocolumn]{revtex4}
\usepackage{graphics}
\usepackage{epsfig}

\begin{document}

\title{Critical behavior of an even offspringed branching and
annihilating random walk cellular automaton with spatial disorder}

\author{G\'eza \'Odor (1) and N\'ora Menyh\'ard (2)}

\affiliation{(1) Research Institute for Technical Physics and
  Materials Science, \\
(2) Research Institute for Solid State Physics,
H-1525 Budapest, P.O.Box 49, Hungary}    

\begin{abstract}
A stochastic cellular automaton exhibiting parity conserving class
transition has been investigated in the presence of quenched 
spatial disorder by large scale simulations. Numerical evidence has been
found that weak disorder causes irrelevant perturbation for the
universal behavior of the transition and the absorbing phase of this
model. This opens up the possibility for experimental observation of 
the critical behavior of a nonequilibrium phase transition to
absorbing state. For very strong disorder the model breaks up to
blocks with exponential size distribution and continuously changing
critical exponents are observed. For strong disorder the randomly
distributed diffusion walls introduce another transition within the 
inactive phase of the model, in which residual particles survive the 
extinction. The critical dynamical behavior of this transition has
been explored.
\end{abstract}
\maketitle

\section{Introduction}

The classification of the universality classes of nonequilibrium phase
transitions to absorbing states is still an open problem of
statistical physics \cite{Dick-Mar,Hin2000,dok}. Reaction-diffusion
(RD) models exhibiting phase transitions to absorbing states bear with 
a particular interest since many other types of systems like surface
growth, spin systems or stochastic cellular automata can be mapped on them.
Unfortunately there hasn't been experimental verification of such
classes except the coagulating random walk: $A+A\to A$ (CRW) 
in one dimension \cite{1dexp}. This is mainly due to the fact that 
the most well known, robust directed percolation (DP) class
\cite{DP} is sensitive to disorder \cite{Janssen97b,CGM98,HIV04,VD05}, 
which occurs in real systems naturally. 
It would be very important to find some other nonequilibrium class, 
which proves to be less sensitive to disorder, hence would 
provide a candidate for experimental verification.

The study  of disordered systems is a hot topic of current research of
statistical physics \cite{overview}. A principal condition for the
relevancy of disorder is the Harris criterion \cite{Harris,Noest} set
up for equilibrium systems and has been found to be valid in some
nonequilibrium models. According to this criterion the pure critical 
point is stable against disorder if the spatial correlation length 
critical exponent $\nu_{\perp}$ fulfills the inequality
\begin{equation} \label{Harriscrit}
d \nu_{\perp} > 2,
\end{equation}
where $d$ is the spatial dimensionality. However an exception is 
reported very recently \cite{jensentempq} for DP with temporal
disorder. Note that for CRW (which is exhibits the same scaling
behavior as the $A+A\to\emptyset$ annihilating random walk (ARW) in
1d) this criterion predicts relevant spatial disorder, 
($\nu_{\perp}=1$) still experiment \cite{1dexp} did not report
measurable effect of randomness unless very strong 
disorder fractures the medium.

Besides the robust DP an other well known universality class is
the so called ``parity conserving'' (PC) class of 1d nonequilibrium
transitions. This was discovered in a one-dimensional stochastic 
cellular automata (CA) exhibiting $Z_2$ symmetric absorbing states 
and domain walls following even offspringed branching and 
annihilating random walk: $A\to 3A$, $2A\to\emptyset$ (BARW2)
\cite{Gras84}. Later it was observed by numerical studies of other models 
\cite{Taka,IJen94,nekim,Park94,2-BAW,meorcikk,Bas,Hin97,ZAM03,GMFPCcikk} 
and field theoretical studies \cite{Cardy-Tauber,HCDMlett} 
confirmed the existence of a corresponding fixed point 
distinct from that of DP. For a review see \cite{MeOdof}. 
This class is also called as directed Ising, DP2 or generalized 
voter model class.

According to the Harris criterion disorder should be relevant for the
critical behavior of this class ($\nu_{\perp}=1.857(1)$\cite{dok}). 
In contrast to this a recent renormalization group (RG) study \cite{HIV04} 
did not find a strong disorder fixed point like in case of DP. 
The question naturally arises if BARW2 is really insensitive to
disorder or the RG method \cite{HIV04} is not applicable for this case.
The principal aim of the present study is to answer this
question. Additionally in the absorbing phase of the BARW2 model the 
ARW dynamics dominates, which has also been addressed in the studies 
\cite{SM98,DM99}. The renormalization study of ARW with spatial
randomness in the reaction rates found marginal perturbations to the 
fixed point of the pure system \cite{DM99}. On the other hand an 
exact study of the infinite reaction rate ARW with space-dependent 
hopping rates found non-universal power-law decay of the density 
of A-s below a critical temperature \cite{SM98}.

Note that in \cite{HIV04} the strong disorder is defined in such a way 
that it can not completely block reactions or diffusion of the
reactants. Therefore the so called infinitely strong fixed point of
\cite{HIV04} does not correspond to the blocking case. 
Such blocking or complete dilution was studied in a 1d toy model 
of random quantum ferromagnetic Ising model \cite{F99} where
continuously variable power laws were found at the phase transition
point. The effect of disconnected domains in the reactions of CRW and 
ARW has been investigated in \cite{MA00}. This study reported stretched
exponential decay in case of exponential domain size distributions and
continuously changing density decay for blocks distributed in a
power-law manner. In the 1d model we investigate such complete
blocking may also occur, hence we investigate this topological effect.

\section{The NEKIMCA model}

To study PC class transitions with disorder we have chosen a very
simple stochastic cellular automaton (SCA) the NEKIMCA introduced in 
\cite{meorcikk}.
It is easy to show that the dual variables of spins ($\uparrow$) the
kinks ($\bullet$) exhibit BARW2 dynamics via the synchronous spin-flip 
dynamics. In this SCA parity conserving kink branching is also
generated due to the synchronous spin update of neighboring sites 
without introducing an explicit spin-exchange reaction as in case of
the NEKIM model \cite{nekim}. The reactions are like the followings:
\begin{itemize}
\item Random walk :  
$\uparrow \ \ \uparrow\bullet\downarrow \ \stackrel{w_i}{\longrightarrow} \
  \uparrow\bullet\downarrow \ \ \downarrow$
\item Annihilation :
$\uparrow\bullet\downarrow\bullet\uparrow \
  \stackrel{w_o}{\longrightarrow} \   \uparrow \ \ \uparrow \ \ \uparrow$
\item Branching :
$\uparrow \ \ \uparrow\bullet\downarrow \ \ \downarrow   \  \stackrel{w_i^2}{\longrightarrow} \  \uparrow\bullet\downarrow\bullet\uparrow\bullet\downarrow$
\end{itemize}
In the NEKIMCA there are two independent parameters parametrized as
\begin{eqnarray}
w_i=\Gamma (1-\delta)/2 \\ w_o=\Gamma (1+\delta) 
\end{eqnarray}

In the computer the state of a single spin is represented by a 1 or 0
of a 32 or 64 bit word $s(j)$ (depending on the CPU type). Hence 32 or
64 CA samples (exhibiting different random initial conditions but the
same quenched noise) updated at once.

The following bit-parallel algorithm was used for the update of states
$s(j)$ at site $j$. A random number $x(j)\in(0,1)$ is selected with 
uniform distribution. If 
\begin{equation}
x(j) < q_i(j) = w_i + \chi(j)
\end{equation} 
a spin-flip, corresponding to random walk of the dual variable
\begin{equation} \label{wirea} 
s'(j) = [ s(j+1) \land s_(j-1) ] \land s(j)
\end{equation} 
is written to all bits of $s(j)^,$.
Following this another uniformly distributed random number
$y(j)\in(0,1)$ is chosen and if 
\begin{equation}
y(j) < q_o(j) = w_o + \chi(j)
\end{equation}
a spin-flip, corresponding to annihilation of the dual variables 
\begin{equation} \label{worea} 
s'(j) = ( [s(j-1) \land s(j)] \ \& \ [s(j+1) \land \ s(j)]  ) \ \land s(j)
\end{equation}
is performed. Here $\chi(j)$ denotes the quenched random noise variable
with uniform distribution 
\begin{equation}
\chi(j) \in (-\epsilon,\epsilon) \ ,
\end{equation}
$\land$ and $\&$ are the logical XOR and AND of computer words.
Note that for strong disorder $q_i(j)$ or $q_o(j)$ may be less than
$0$ at a site meaning a blockade for that reaction.

A single Monte Carlo step (MCS) consists of updating all $s'(j)$
sites with periodic boundary conditions and writing back $s(j)=s'(j)$ 
for $j\in(0,L-1)$ (throughout the paper the time is measured by MCS).

\section{Dynamical simulations along the disordered PC transition
  line} \label{1dsimu}

The simulations were carried out on $L= 4\times 10^4 - 10^5$ sized systems.
with periodic boundary conditions. The initial states were randomly
half filled lattices, and the density of kinks is followed up
to $10^7 - 10^8$ MCS. We started the exploration of the $(\delta, \epsilon)$
phase diagram by determining the phase transition of the impurity free 
($\epsilon=0$) case with $\Gamma=1$. This was located at
$\delta_c=-0.550(1)$ as a power-law decay of the kink density
\begin{equation}
\rho\propto t^{-\alpha}
\end{equation}
with exponent $\alpha=0.28(1)$. This is in good agreement with the PC
class value \cite{meorcikk}. 
\begin{figure}
\epsfxsize=70mm
\epsffile{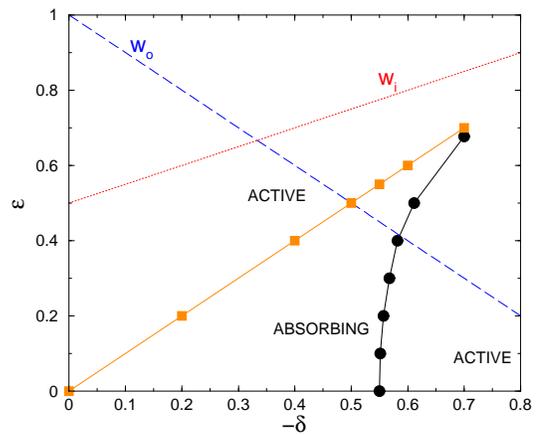}
\vspace{2mm}
\caption{(Color online) Phase diagram of the disordered NEKIMCA for negative
$\delta$. Bullets correspond to the disordered PC class transition points,
squares to the freezing transition (lines are for the guide of eye
only). The dashed line shows $w_o$, while dotted line denotes
$w_i$. \label{deltac}}
\end{figure}
Next we introduced the quenched disorder and determined $\delta_c$ 
for different $\epsilon$ values.
Table \ref{PCT} and Fig.\ref{deltac} show the results. 
\begin{table}[h] 
\caption{Numerical results for the disordered PC class transition line.
\label{PCT}}
\begin{tabular}{|l|l|l|l|l|l|}
\hline
 $\epsilon$ & $-\delta_c$ & $\alpha$ & $\beta$  & $z$ & $\eta$\\
\hline
$0.0$ & $0.550(1)$ & $0.280(6)$ &    &  & \\
$0.1$ & $0.5513(5)$& $0.280(6)$ &    &  & \\
$0.2$ & $0.557(1)$ & $0.280(6)$ & $ $ & & \\
$0.3$ & $0.5676(1)$& $0.280(5)$ & $0.95(1) $ & $1.11(1)$ & $0.285(5)$ \\
$0.4$ & $0.5849(1)$& $0.265(10)$ &     &  &\\
$0.5$ & $0.6115(4)$& $0.25(1)$  & $0.84(2)$ & $1.0(1) $ & $0.252(6)$  \\
$0.6767(2)$& $0.7$ & $0.22(1)$  & $0.80(2)$ &  & \\
\hline
\end{tabular}
\end{table}
As one can see the disorder moves the transition point in such a way
that the size of absorbing phase increases along the $-\delta$ axis.

\begin{figure}
\epsfxsize=70mm
\epsffile{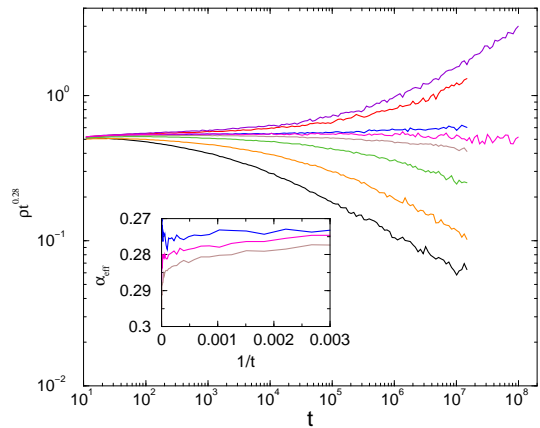}
\vspace{2mm}
\caption{(Color online) Density decay of NEKIMCA for $\epsilon=0.3$, $\Gamma=1$ for
 different $\delta=0.571,0.57,0.568$,$0.5675,0.567,0.0.565,0.56,0.55$ 
(from top to bottom). The insert shows the
 local slopes for $\delta=0.568, 0.5675, 0.567$ (from top to bottom).}
\label{qpc3}
\end{figure}

Weak disorder ($\epsilon < w_o$) seems to be irrelevant (see
Fig.\ref{qpc3}). The line corresponding to $\delta=0.5675$ can well be
described by a power law from $10^3 < t < 10^8$ MCS (5 decades !)
with exponent $\alpha=0.28$.
The local slopes (effective exponents) of the kink density decay are defined as
\begin{equation}
\alpha_{eff}(t) = {- \ln \left[ \rho(t) / \rho(t/m) \right] 
\over \ln(m)} \label{slopes}
\end{equation}
(where we used $m=2$). At the critical point the $\alpha_{eff}(t)$ curve
exhibits a straight line shape for $t\to\infty$, while in 
sub(super)-critical cases $\alpha_{eff}(t)$ curves veer down(up) 
respectively. One can read off $\alpha_{eff}\to\alpha = 0.28(1)$ in a
perfect agreement with the PC class value.
Similar results are obtained for $\epsilon=0.1, 0.2, 0.4$ and
summarized in Table \ref{PCT}.

\begin{figure}
\epsfxsize=70mm
\epsffile{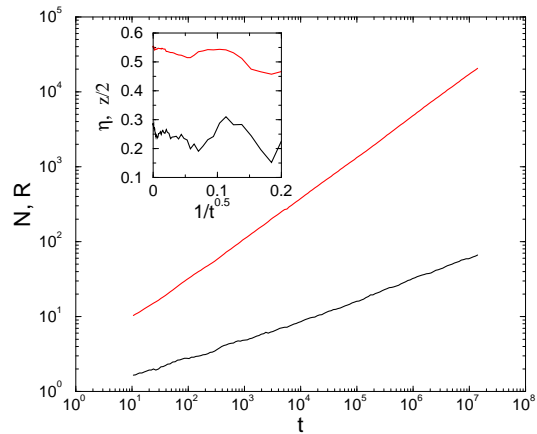}
\vspace{2mm}
\caption{(Color online) Spreading in NEKIMCA for $\epsilon=0.3$, $\Gamma=1$
  $\delta_c=-0.5679$. The insert shows the corresponding local slopes.}
\label{qpcspins_3}
\end{figure}

To confirm these results we run cluster spreading simulations from a
single active seed (one kink, odd sector) and measured the number 
of particles ($N(t)$) and the diameter of clusters ($R(t)$).
At the critical point these are expected to scale as
\begin{equation}
N(t) \propto t^{\eta} \ , R(t) \propto t^{z/2}
\end{equation}
As Figure \ref{qpcspins_3} shows both quantities exhibit power-law
behavior with PC class exponents: $\eta = 0.285(5)$, $z=1.11(1)$
\cite{IJen94,MeOdof}.

\begin{figure}
\epsfxsize=70mm
\epsffile{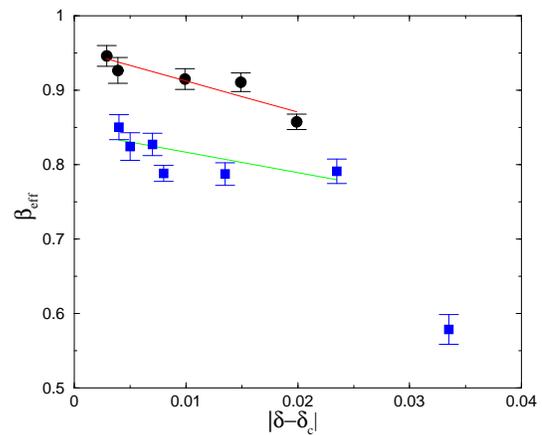}
\vspace{4mm}
\caption{(Color online) Effective order parameter exponent results on the disordered
PC class line for $\epsilon=0.3, 0.5$ (top to bottom). Lines correspond to
linear fit.}
\label{beta_3}
\end{figure}

Finally we performed steady state simulations in the active phase
at $\epsilon=0.3$. We followed the density of kinks in many
realizations until saturation is reached and averaged in a time window 
following that point.
The steady state density in the active phase at a critical phase
transition is expected to scale as
\begin{equation}
\rho(\infty,\delta) \propto |\delta-\delta_c|^{\beta} \ .
\end{equation}
Using the local slopes method one can get a precise estimate for
$\beta$ as well as for the corrections to scaling
\begin{equation}
\beta_{eff}(\delta_i) = \frac {\ln \rho(\infty,\delta_i) -
\ln \rho(\infty,\delta_{i-1})} {\ln(\delta_i) - \ln(\delta_{i-1})} \ \ ,
\label{beff}
\end{equation}
where we used the $\delta_c$ value determined before. The local slopes
analysis resulted in $\beta_{eff}\to\beta = 0.95(1)$ in agreement 
with the PC class value again \cite{GMFPCcikk} (see Fig.\ref{beta_3}).

\subsection{Annihilation blocking on the disordered PC transition line}

In the inactive phase and at the critical point depletion of kinks
dominates via the $AA\to\emptyset$ annihilation. 
However for strong enough disorder active domains of arbitrary 
sizes may also emerge due to the exponential distribution of such
events. The contribution of these such large regions can be estimated
as in \cite{Noest} with the difference that for PC class even the
absorbing phase decays algebraically i.e. one does not have
exponential decay that could slow down to stretched exponential at
some ``clean critical point'' as in the DP case.
The probability $p_a$ for finding a rare region 
of size $l_a$ is
\begin{equation}
p_a \propto \exp(-q l_a) \ ,
\end{equation}
where $q \propto w_i^2 - w_o + \epsilon$ is the probability that a site
is active. The long-time kink decay density is dominated by these rare 
regions. For long times any finite region decays exponentially hence  
\begin{equation}
\rho(t) \propto \int dl_a l_a p_a \exp(- t/\tau(l_a) \  , \label{Ndec}
\end{equation}
where $\tau(l_a)$ is the characteristic decay time of a size of
region of size $l_a$. The average lifetime of such active regions
grows as 
\begin{equation}
\tau(l_a) \propto \exp(a l_a)
\end{equation}
because a coordinated fluctuation of the entire region is required to
take it to the absorbing state. The saddle point analysis of (\ref{Ndec})
as in \cite{Noest} results in power-law decay
\begin{equation}
\rho(t) \propto t^{-p_a/a} \ .
\end{equation}
Hence one may expect continuously changing decay exponents either as
the effect of rare active regions or as the effect of high diffusion
barriers \cite{SM98}.

As we discussed in the previous section we did not see the occurrence
of such continuous decay variables for weak disorder. However by going
above $\epsilon >~\sim 0.4$ the situation changes and we begin to see
deviation from the pure PC decay behavior (see Table \ref{PCT})
The effect of disorder becomes drastic. 
By increasing $\epsilon$ above $w_o$ the $q_o$ reaction probabilities 
may become zero at certain sites randomly (with probability $p_w$) and 
the kink annihilation is blocked at those sites. From annihilation
reaction point of view the systems falls apart to $l$ blocks with an 
exponential probability distribution
\begin{equation} 
p(l) =  p_w p_{nw}^l  \ , \label{ldist}
\end{equation}
where the no-wall probability $p_{nw}$ is related to $\epsilon$ 
according to our algorithm as $p_{nw} = 1 - p_w = 1 - (\epsilon-w_o)$, 
i.e where $q_o < 0$ is set.
This means that neighboring kinks can't even annihilate at the block
boundary points which may cause relevant 
perturbation to the critical behavior.

\begin{figure}
\epsfxsize=70mm
\epsffile{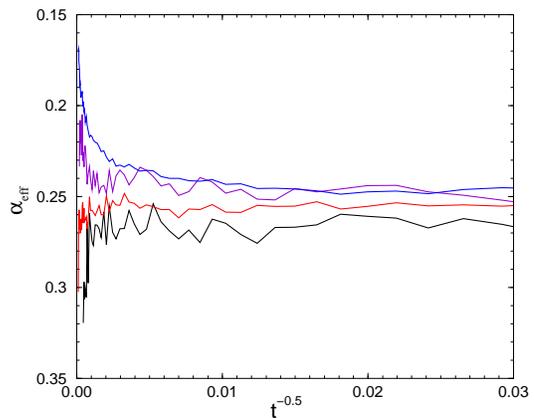}
\vspace{4mm}
\caption{(Color online) Effective $\alpha$ in the strong disorder region of the
  disordered PC class transition line ($\epsilon=0.5$), 
$-\delta_c = 0.613, 0.612, 0.611, 0.609$ (top to bottom).
}
\label{qpc_5}
\end{figure}

As Figure \ref{qpc_5} shows for $\epsilon=0.5$ the phase transition 
occurs at $\delta_c=-0.6115(3)$ and the density decay exponent is
smaller than the PC class value: $\alpha = 0.25(1)$.
For $\delta=-0.7$, $\epsilon_c=0.677(1)$, it is even smaller: $\alpha
= 0.22(1)$ suggesting continuously changing exponents along this
transition line by increasing $\epsilon$. 
Similarly the steady state exponent $\beta$ and
the spreading exponent $z$ changes in this region 
(see Table \ref{PCT} and Fig.\ref{beff}).

\section{Dynamical simulations in the inactive phase} \label{Inact}

\begin{figure}
\epsfxsize=70mm
\epsffile{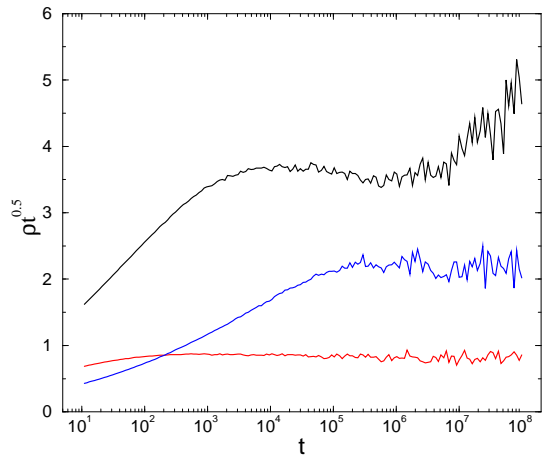}
\vspace{4mm}
\caption{(Color online) Density decay in the inactive phase at coordinates:
($\delta=-0.6$, $\epsilon=0.57$),($\delta=-0.5$, $\epsilon=0.3$),
($\delta=-0.525$, $\epsilon=0.1$) top to bottom).}
\label{ina1-5}
\end{figure}

In the inactive phase annihilating random walk dominates, in which for
pure systems the density decays asymptotically as a power-law \cite{Lee}
\begin{equation} \label{sqrtlaw}
\rho\propto t^{-1/2} \ ,
\end{equation}
hence one can check the effect of disorder for this process. 
As one can see on Fig.\ref{ina1-5} the disorder does not change this
behavior for small $\epsilon$, the $\rho(t)t^{1/2}$ curves level off.  
For larger $\epsilon$ however the deviations from the (\ref{sqrtlaw}) 
law is observable for long times. Following a transient region, which
is faster than (\ref{sqrtlaw}) and characteristic of ARW with finite 
reaction rates \cite{BMGR92} the decay slows down. 
For example for $\epsilon=0.57$, $\delta=-0.6$ power-law fitting for 
$t > 10^6$ MCS results in $\alpha=0.43(1)$, while for $\epsilon=0.48$, 
$\delta=-0.5$ we obtained $\alpha=0.46(1)$. These $\epsilon$ dependent 
$\alpha$ values are in agreement with the results of \cite{SM98},
where continuously changing power-law exponents are determined below a 
critical temperature in case of exponentially distributed diffusion 
barrier heights.

\section{Dynamical simulations of the freezing transition} \label{freez}

As one can see on Figure \ref{deltac} for negative $\delta$ values
the PC class transition line does not cross the $w_i$ line meaning
that the diffusion and branching reactions are not blocked. However
due to the parallel computer algorithm we used in case of strong
disorder ($\epsilon \ge 1-w_o$) $q_o$ may become 1 at certain sites 
hence a random walk move is immediately overwritten by the
(\ref{worea}) CA update at these sites. 
As the consequence in the absorbing phase (above the disordered PC transition
line) the diffusion is blocked. Again the system falls apart to blocks 
with closed diffusion boundaries. As the depletion goes on kinks can
not leave the blocks to interact with others in neighboring ones 
and ``frozen'' steady states emerge for $\epsilon > -\delta$. 
Note however that the system is not completely frozen,
kinks can not only diffuse within the blocks, but branching may occur with 
(small probability) followed by a quick annihilation. In any case the
residual density remains small.

\begin{figure}
\epsfxsize=70mm
\epsffile{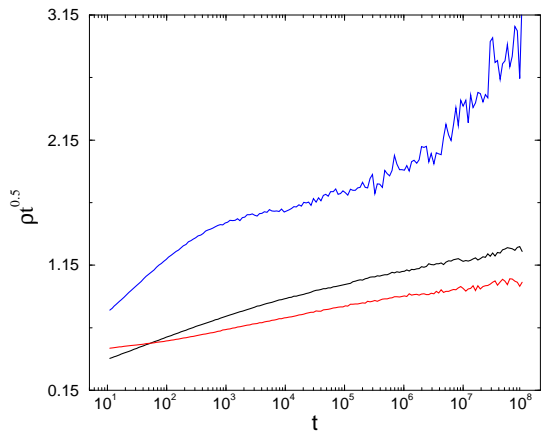}
\vspace{4mm}
\caption{(Color online) Density decay on the freezing transition line for
  $\epsilon=0.6,0.4,0.2$ (top to bottom).}
\label{negdeltadec}
\end{figure}

We run dynamical simulations on the freezing transition line and as 
Figure \ref{negdeltadec} shows the density decay is slower than the 
square-root law of ARW (\ref{sqrtlaw})) and changes continuously along
the freezing transition line (see Table \ref{freezT}). However for 
$\epsilon \le 0.5$ this deviation can be interpreted as a
(logarithmic) correction to scaling.
For $\epsilon > 0.5$ even the $w_o$ (annihilation) reactions are
blocked and the deviation from (\ref{sqrtlaw}) is strong. We repeated 
this simulation on a system with different size ($L=10^4$), but did 
not see any size dependence.  
\begin{table}[h] 
\caption{Numerical results for the freezing transition line.
\label{freezT}}
\begin{tabular}{|l|l|l|l|l|l|}
\hline
$\epsilon$&$-\delta_c$&$\alpha$ & $\beta$  & $z$ & $N(\infty)$\\
\hline
$0.2$ &  $0.2$     & $0.47(3)$ & $1.00(5)$ & $0.95(5)$ & $1.031(1)$ \\
$0.4$ &  $0.4$     & $0.47(3)$ &          & $0.95(5) $ & $1.275(5)$ \\
$0.5$ &  $0.5$     & $0.43(1)$ &          & $1.0(2)$ &   $1.85(5)$  \\
$0.6$ &  $0.6$     & $0.41(1)$ & $0.75(3)$& $0.76(2)$ &  $3.75(5)$ \\
$0.7$ &  $0.7$     & $0.39(1)$ & $0.68(4)$& $0.72(2)$ &  $14(1)$  \\
\hline
\end{tabular}
\end{table}

\begin{figure}
\epsfxsize=70mm
\epsffile{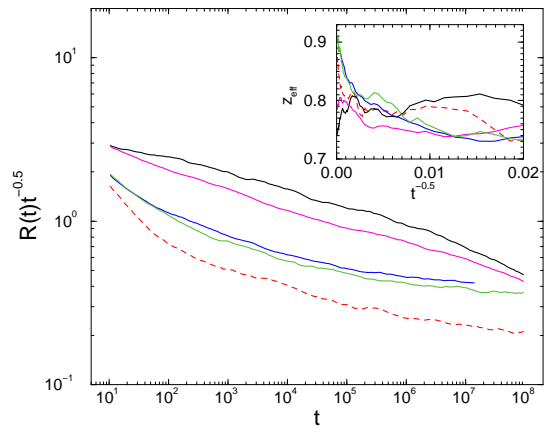}
\vspace{4mm}
\caption{(Color online) Spreading in NEKIMCA along the freezing transition line
  $\epsilon=0.7,0.6,0.0,0.2,0.125$ (top to bottom). The 
$\epsilon=0,0.125$ data correspond to $\Gamma=0.5$ the others to
  $\Gamma=1$. The $\epsilon=0.125$ line corresponds to $\delta=0.5$
  the others to $\epsilon=-\delta$. The insert shows the corresponding
  local slopes.}
\label{spreadins}
\end{figure}

The spreading exponent $z$ changes similarly along the freezing
transition line. It is close to $z=1$ for $\epsilon \le 0.5$ and
deviates it significantly for $\epsilon > 0.5$ 
(see Figure \ref{spreadins}).
The average number of kinks ($N(t)$) evolves to a small constant value as 
$t\to\infty$ ($\eta=0$) all along the freezing transition line. This value 
changes continuously by increasing the disorder strength. For
$\epsilon \le 0.2$ only a single kink survives ($N(\infty) \simeq 1$), while
for other values see Table \ref{freezT}.
 
\begin{figure}
\epsfxsize=70mm
\epsffile{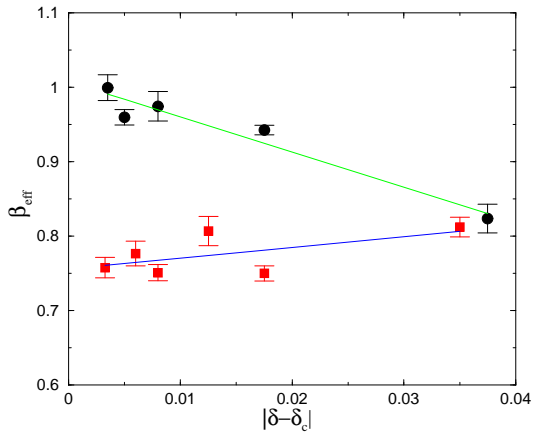}
\vspace{4mm}
\caption{(Color online) Effective order parameter exponent results on the freezing transition
line for $\epsilon=0.2, 0.6$ (top to bottom). Lines correspond to
linear fit.}
\label{betafr}
\end{figure}

In the frozen (active) phase in the long time limit pairs within the
same block annihilate and basically lonely kinks are wandering in confined
regions of sizes $l$ (local $A\to 3A\to A$ reactions are also permitted
with a small probability). Therefore the density of frozen kinks depend on
the density of blocks, hence on $\epsilon$. The concentration of blocks
$c_b$ can be expressed with the average length $<l>$ of blocks as
\begin{equation}
\rho \propto c_b \propto <l>^{-1}  \ .
\end{equation}
For the exponential block size distribution (\ref{ldist}) the average
block size is
\begin{equation}
<l> =  (-\ln(p_{nw}))^{-1}
\end{equation}
with the no-wall probability 
\begin{equation}
p_{nw}= 1 - p_w = 1 - (w_o+\epsilon-1) = 2-\epsilon-w_o \ .
\end{equation}
The kink density is  
\begin{equation}
\rho \propto -\ln(2-\epsilon-w_o)  \ ,
\end{equation}
which for small wall probability $p_w$ has the leading order singularity
\begin{equation}
\rho \propto p_W \ .
\end{equation}
This means that in the frozen phase $\beta=1$. The simulations
confirm this. For $\epsilon < 0.5$, where only diffusion traps are present 
the $\beta\sim 1$ indeed (see Table \ref{freezT} and Figure \ref{betafr}).
For $\epsilon \ge 0.5$ where annihilation is also
blocked the exponent $\beta$ decreases by increasing $\epsilon$ 
(see Table \ref{freezT}).

\begin{figure}
\epsfxsize=70mm
\epsffile{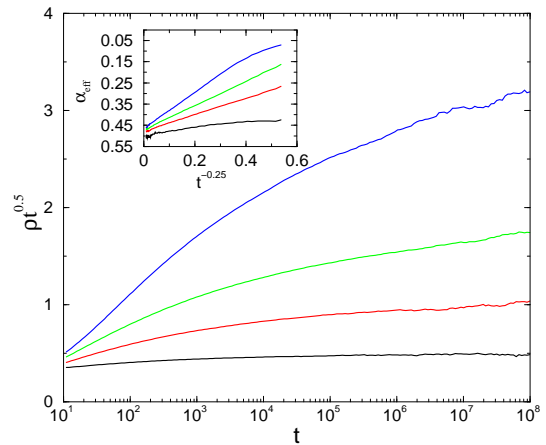}
\vspace{4mm}
\caption{(Color online) Density decay in NEKIMCA for positive $\delta$
along the freezing transition line
  $\epsilon=0.075,0.125,0.175,0.25$ (top to bottom). 
  The insert shows the corresponding local slopes.}
\label{posdel}
\end{figure}

We have also investigated the phase space for positive $\delta$ 
(with $\Gamma=0.5$) where $\epsilon > w_i$ causes a direct
freezing transition via diffusion traps in the absorbing phase.
The density decay simulations (see Figure \ref{posdel}) show small
deviations from the square-root law of ARW (\ref{sqrtlaw})), which can
be interpreted as corrections to scaling.

\section{Discussion} \label{disc}
We have investigated the phase diagram of the NEKIMCA as the function
of quenched disorder with large scale simulations. We have chosen
NEKIMCA, a very simple SCA, which is known to exhibit PC class
transition of kinks to absorbing states. We determined static and 
dynamic exponents ($\alpha$,$\beta$,$\eta$,$z$) at several
points along the phase transition line and in the inactive phase. 
The main consequence of this survey is that weak disorder does not 
change the scaling behavior of the quantities studied. 
This supports the results of a recent real space RG study \cite{HIV04} 
but contradicts the Harris criterion. 

A possible resolution of this contradiction could be the study of the 
distribution of the critical point coordinates over the ensemble of 
samples of size $L$.
According to recent progresses in the finite size scaling theory of
disordered systems, thermodynamic observable are not self-averaging
at critical points when the disorder is relevant in the
Harris criterion sense  \cite{domany95,AH,Paz1,domany,AHW,Paz2}. 
This lack of self-averageness at criticality is directly related to 
the distribution of pseudo-critical temperatures $T_c(i,L)$ over the 
ensemble of samples $(i)$ of size $L$. 
This has been shown numerically in case of a wetting transition of a
polymer chain with quenched disorder \cite{MG0509}. An interesting
further direction of research would be the study of those
distributions for nonequilibrium systems like for NEKIMCA.

For weak disorder ($\epsilon < \simeq 0.4$) we did not see deviations 
from the asymptotic square root power-law (\ref{sqrtlaw}) in the 
inactive phase. However for stronger disorder continuously changing 
density decay exponent ($\alpha$) were observed. 
This corroborates a former RG study \cite{DM99} for ARW with site 
dependent reaction rates and an exact calculation for the infinite 
reaction rate ARW with disordered diffusion traps \cite{SM98}.

Very strong disorder in our model may introduce complete blocking of 
reactions or diffusion with exponential domain size distribution.
If the system becomes segmented by diffusion walls a freezing
transition to fluctuating states occur. In this case in odd parity 
blocks residual particles remain active. Large scale simulations
suggest that this kind of disorder is marginal, the density decay
slows down from the (\ref{sqrtlaw}) power-law by a logarithmic
correction or changes continuously if the annihilation is blockaded too.
In the frozen phase the concentration of residual kinks
increases logarithmically, with $\beta_{eff}\to 1$ asymptotically. 
This is in agreement with the results of \cite{MA00}, who considered 
the ARW model with complete blockades. However we did not see
crossover to stretched exponential decay reported in \cite{MA00}
for long times, but rather the decay slows down in our model. 
This is due to the fact that in our model in the frozen state not only
blockades exist, but reaction and diffusion probabilities are
randomized, which can introduce slower power-laws 
\cite{Noest,SM98,DM99}.

In the region of the phase diagram where the disorder is so strong 
that the annihilation reaction becomes segmented we found continuously 
changing exponents along the disordered PC and the freezing 
transition lines both.

\vskip 0.5cm

\noindent
{\bf Acknowledgements:}\\
We thank F. Igl\'oi and G. M. Sch\"utz for the useful comments.
Support from Hungarian research funds OTKA (Grant No. T046129) is acknowledged.
The authors thank the access to the NIIFI Cluster-GRID, LCG-GRID 
and to the Supercomputer Center of Hungary.

\end{document}